\documentclass[%
 reprint,
 amsmath,amssymb,
 aps,
]{revtex4-2}

\usepackage{graphicx}
\usepackage{dcolumn}
\usepackage{bm}

\usepackage{xcolor}
\begin{document}

\preprint{APS/123-QED}

\title{Inverse Clausius Thermodynamics in Run-and-Tumble Dynamics}

\author{Oded Farago}
\affiliation{%
 Department of Biomedical Engineering, Ben-Gurion University of the Negev, Be’er Sheva 85105, Israel
}%

\date{\today}

\begin{abstract}

We establish a mapping between one-dimensional run-and-tumble particle dynamics in the presence of thermal noise, and overdamped Brownian motion in a spatially inhomogeneous temperature field. The approach is formulated as an inverse-Clausius thermodynamic framework, where the effective temperature is inferred from the steady-state density. Within this mapping, the local entropy flux and entropy production rate can be extracted directly from steady-state observables, without requiring explicit knowledge of the full position–velocity distribution. The framework introduces a closure that, to leading order, relates entropy flow to spatial variations of effective temperature, yielding a picture of entropy transfer from hotter to colder regions. We apply the approach to two-state and multistate run-and-tumble models in harmonic and nonlinear potentials. For harmonic confinement, the closure is exact in the two-state case and very accurate in multistate models. For nonlinear potentials that are locally confining (positive curvature at the origin), comparable accuracy is observed. In contrast, potentials with vanishing or negative curvature require higher-order corrections and reveal a correspondence between the spatial structure of the entropy production rate and that of the confining potential.

\end{abstract}

\maketitle


Run-and-tumble dynamics is a widely used model for self-propelled particles. It was originally introduced to describe bacterial motion such as \textit{E. coli}~\cite{berg1972chemotaxis}, and later adopted in statistical physics as a framework for studying active matter~\cite{tailleur2008statistical,khatami2016active,sandor2017collective,gutierrez2021collective,kurzthaler2024characterization,roberts2023ratchet}. Its appeal lies in its simplicity, ability to represent key non-equilibrium features, and analytical tractability. A one-dimensional run-and-tumble particle (RTP) model describes a particle evolving in a force field $F(x)=-\partial U/\partial x$ while carrying an ``active'' velocity $v$ that is randomized at tumbling events. Denoting by $P(x,v,t)$ the joint probability density, and by $P(v)$ the {\em symmetric} distribution function from which $v$ is assigned at a tumble, the dynamics is governed by the Fokker--Planck (FP) equation
\begin{eqnarray}
&&\!\!\!\!\!\!\!\!\!\frac{\partial P(x,v,t)}{\partial t} =
-\frac{\partial}{\partial x}\Big[(\mu F(x)+v)\,P(x,v,t)
-D\,\frac{\partial P(x,v,t)}{\partial x}\Big]  \nonumber\\
&&\qquad -\,r\big[\,P(x,v,t)-\,P(x,t)\,P(v)\big],
\label{eq:FPmodified}
\end{eqnarray}
where $P(x,t)=\int_{-\infty}^{\infty} P(x,v,t)\,dv$ is the position marginal distribution. Here $\mu$ is the mobility, and $D=\mu k_B T$ is the diffusion coefficient. The first term on the right-hand side of Eq.~(\ref{eq:FPmodified}) can be written as $-\partial_x J(x,v,t)$, where
\begin{equation}
J(x,v,t) = (\mu F(x) + v)\, P(x,v,t) - D\, \frac{\partial P(x,v,t)}{\partial x}
\label{eq:partialflux}
\end{equation}
is the partial flux for particles with velocity $v$ at position $x$.
The second term describes tumbling events at rate~$r$. The standard~two-state RTP (s-RTP) is a Markov process in which the velocity switches between $\pm v_0$, with exponentially distributed run times. In this case we use $r/2$ as the switching rate, since a tumble flips the velocity rather than redraws it.
Introducing $p_\pm(x,t)\equiv P(x,v=\pm v_0,t)$, 
we write
\begin{align}
\frac{\partial p_+}{\partial t}
&= -\frac{\partial}{\partial x}\!\left[\big(\mu F(x)+v_0\big)p_+\right]
+ D\,\frac{\partial^2 p_+}{\partial x^2}
-\frac{r}{2}p_+ + \frac{r}{2}p_-, \nonumber\\
\frac{\partial p_-}{\partial t}
&= -\frac{\partial}{\partial x}\!\left[\big(\mu F(x)-v_0\big)p_-\right]
+ D\,\frac{\partial^2 p_-}{\partial x^2}
-\frac{r}{2}p_- + \frac{r}{2}p_+ .
\label{eq:FPstandard}
\end{align}

In the absence of thermal noise ($D=0$), analytical steady-state solutions exist for arbitrary confining forces $F(x)$, yielding a density $P(x)=p_+(x)+p_-(x)$ with compact support determined by $|\mu F(x)|\le v_0$~\cite{dhar2019}. For $D\neq0$, analytical solutions for the steady-state distribution (SSD) are limited to specific cases, such as harmonic~\cite{frydel2022run} or piecewise-linear~\cite{singh2023} traps. Improtantly, even weak thermal noise ($D\ll D_{\rm ac}=v_0^2/r$, where $D_{\rm ac}$ is the non-thermal s-RTP diffusion coefficient) qualitatively changes the steady state by rendering the dynamics ergodic over the full one-dimensional space. This sensitivity to thermal noise motivates the alternative description developed below, which extracts local thermodynamic information directly from steady-state observables without requiring explicit solutions of Eq.~(\ref{eq:FPstandard}).

Effective-temperature ideas provide a useful framework in this context, bridging
statistical-mechanical modeling of active systems and thermodynamic
perspectives~\cite{Bowick2022,Jager2024}. Related approaches include phase
equilibrium descriptions of scalar active matter in the context of
motility-induced phase separation~\cite{Solon2018,Maggi21}, fluctuation--dissipation
based effective temperatures in active and driven systems~\cite{Loi2008,Cugliandolo2011},
and generalized thermodynamic frameworks restoring Clausius- and Carnot-type
bounds in systems violating the Einstein relation~\cite{Sorkin2024}. Of particular
relevance here are mappings of active dynamics onto auxiliary descriptions with
spatially varying temperature profiles~\cite{Mandal2019,horwitz19}.

Entropy production is a key measure of irreversibility in nonequilibrium systems. A local description is given by the entropy flux density $\phi(x)$ and the entropy production rate density $\pi(x)$, corresponding to entropy dissipation to the environment and local generation in the system, respectively. In steady state, the entropy of the system does not change, so that the total entropy production rate $\dot{S}$ equals the total entropy dissipation to the environment: $\dot{S}=\int \pi(x)\,dx=\int \phi(x)\,dx$.
Their difference defines the net entropy production rate density,
\begin{equation}
\sigma(x)\equiv \pi(x)-\phi(x)=-\frac{d j_s(x)}{dx},
\label{eq:entropybalance}
\end{equation}
where the first equality is a definition and the second expresses a local continuity equation for entropy, in which production and dissipation are balanced by spatial transport via the entropy current $j_s(x)$. Its spatial integral vanishes in steady state, as a consequence of the global entropy balance and its form as a total derivative, with no entropy flux at infinity, so that $j_s(x)\to 0$ as $x\to\pm\infty$.

A central goal of this work is to extract \emph{local} nonequilibrium thermodynamics from steady-state observables accessible in simulations. Langevin dynamics yield the position SSD, $P(x)=\int_{-\infty}^{\infty} P(x,v)\,dv$, and the (normalized) active-flux density (AFD) $\Delta(x)$, defined by $v_0\,\Delta(x)=\int_{-\infty}^{\infty} v\,P(x,v)\,dv$, where $v_0^{\,2}\equiv \int_{-\infty}^{\infty} v^2\,P(0,v)\,dv$ is the local second moment of $v$. Note that $v_0$  can be obtained directly from simulation data by averaging $v^2$ over trajectory segments around $x=0$ without reconstructing $P(0,v)$.
In the two-state s-RTP, where $v=\pm v_0$, $\Delta(x)=p_+(x)-p_-(x)$. The fields $P(x)$ and $\Delta(x)$ already suffice (see below) to determine the entropy flux $\phi(x)$, i.e., the local rate at which entropy is transferred to the thermal bath (see derivation in, e.g.~\cite{paoluzzi2024entropy}),
\begin{equation}
T\,\phi(x)=\frac{1}{\mu}\int_{-\infty}^{\infty} v\,J(x,v)\,dv,
\label{eq:localphi}
\end{equation}
where $J(x,v)$ is defined in Eq.~(\ref{eq:partialflux}). In contrast, the entropy production rate density $\pi(x)$ cannot, in general, be determined from this reduced information alone, as its evaluation requires the full joint distribution $P(x,v)$:
\begin{eqnarray}
\pi(x)&=&k_B \int dv \left\{
\frac{J^2(x,v)}{D\,P(x,v)}\right. \label{eq:kullback-leibler}\\
&+& \left.r\left[P(x,v)-P(x)P(v)\right]
\ln\!\left(\frac{P(x,v)}{P(x)P(v)}\right)
\right\}. \nonumber
\end{eqnarray}

In this Letter, we introduce an inverse--Clausius mapping between RTP dynamics, and overdamped Brownian motion in an inhomogeneous temperature field $T(x)$, directly inferred from the steady-state density $P(x)$. The mapping provides a coarse-grained thermodynamic description that eliminates the need to reconstruct the full joint distribution $P(x,v)$, whose numerical evaluation is costly and error-prone due to high-dimensional sampling and to the numerical differentiation required for computing the currents $J(x,v)$. Within this framework, $\pi(x)$ is approximated in terms of the reduced observables $P(x)$ and $\Delta(x)$. Since the coarse-grained description does not use the full $(x,v)$ statistics, this representation is only approximate. The mapping naturally introduces the temperature deviation field $\theta(x)=T(x)-\langle T\rangle$, which distinguishes regions that are effectively hotter or colder than average, and allows the net entropy production density $\sigma(x)=\pi(x)-\phi(x)$ to be expressed in terms of $\theta(x)$ and $P(x)$. We demonstrate that, to leading order, the net entropy production density $\sigma(x)$ is accurately captured by a local linear relation of the form $\sigma(x)\approx c\,P(x)\theta(x)$, which provides a robust and quantitatively reliable description across a broad range of cases.

We consider an even velocity-resetting distribution $P(v)$ and a symmetric potential $U(x)$, so that the force $F(x)=-\partial_x U(x)$ is odd. Multiplying Eq.~(\ref{eq:FPmodified}) by $v$ and integrating over $v$ yields
\begin{align}
&- \frac{d}{dx} \left[ \mu F(x)\, v_0 \Delta(x) + \langle v^2(x) \rangle P(x) \right] \nonumber \\
&\quad + D\, v_0 \frac{d^2 \Delta(x)}{dx^2} - r\, v_0 \Delta(x) = 0,
\label{eq:FPforDelta}
\end{align}
where $\langle v^2(x) \rangle=[\int_{-\infty}^{\infty} v^2\,P(x,v)\,dv]/P(x)$ is the local mean-squared active velocity (in the s-RTP model, $\langle v^2(x)\rangle = v_0^2$). Integrating Eq.~(\ref{eq:FPforDelta}) with respect to $x$, and using the boundary condition that $P(x)$ and $\Delta(x)$ vanish as $x \to \pm \infty$, yields
\begin{equation}
\mu F(x)\, v_0 \Delta(x) + \langle v^2(x) \rangle P(x)- D\, v_0 \frac{d \Delta(x)}{d x} = - r\, v_0\, I_{\Delta}(x),
\label{eq:FPforI}
\end{equation}
where $I_\Delta(x) = \int_{-\infty}^x \Delta(y)\, dy$ is the primitive function of $\Delta(x)$. We now substitute Eq.~(\ref{eq:partialflux}) for $J(x,v)$ in 
Eq.~(\ref{eq:localphi}) for the local entropy flux $\phi(x)$, and perform the $v$ integration. This gives
\begin{equation}
\phi(x) =  \frac{1}{\mu T}\left[ \mu F(x)\, v_0 \Delta(x) + \langle v^2(x) \rangle P(x)- D v_0 \frac{d \Delta(x)}{dx} \right],
\label{eq:EPRreduced}
\end{equation}
which, by using Eq.~(\ref{eq:FPforI}), simplifies to
\begin{equation}
\phi(x) = -\frac{r v_0}{\mu T} \, I_{\Delta}(x).
\label{eq:EPRreduced2}
\end{equation}
Equation~(\ref{eq:EPRreduced2}) thus provides a simple route for computing $\phi(x)$ from $\Delta(x)$ via its primitive $I_{\Delta}(x)$, obtained by a single numerical integration (e.g., the trapezoidal rule).

This brings us to the evaluation of $\pi(x)$ within the inverse-Clausius framework, which maps the RTP steady state onto overdamped Brownian motion in a position-dependent temperature field $T(x)$. Within this description, the probability current takes the form
$J(x)=\mu\left[F(x)P(x)-\partial_x\big(k_B T(x)P(x)\big)\right]$~\cite{vankampen88,farago19}.
In steady state, $J(x)=0$, reflecting a balance between the deterministic force and the gradient of the effective pressure $k_B T(x)P(x)$. This pressure-gradient term replaces the active and diffusive contributions to the RTP current, $v_0\Delta(x)-\mu k_B T\,\partial_x P(x)$, so that, upon identifying the two alternative forms of the current, one obtains $v_0\Delta(x)-\mu k_B T\,\partial_x P(x) = -\mu\partial_x\!\left[k_B T(x)P(x)\right]$. Integrating this relation
yields $v_0 I_\Delta(x)=\mu k_B P(x)[T-T(x)]$, where $I_\Delta(x)=\int_{-\infty}^x \Delta(y)\,dy$. Using Eq.~(\ref{eq:EPRreduced2}), this gives
\begin{equation}
\phi(x)=\frac{r k_B}{T}\,P(x)\,[T(x)-T].
\label{eq:phi-t}
\end{equation}
Equation~(\ref{eq:phi-t}) admits a direct interpretation as an entropy flux from the system to the bath, $\phi(x)=\dot Q_{\rm out}(x)/T$, with $\dot Q_{\rm out}(x)=r k_B P(x)[T(x)-T]$. This implies that tumbling events, occurring at rate $r$ and distributed in space according to $P(x)$, are associated with a local heat release $Q_{\rm out}=k_B[T(x)-T]$. In this sense, the construction is inverse--Clausius, with the effective temperature $T(x)$ inferred from the entropy flux $\phi(x)$ rather than the other way around.


Within this thermodynamic framework, we seek a relation between the entropy production rate $\pi(x)$ and the effective temperature field $T(x)$, in analogy with the expression obtained for the entropy flux $\phi(x)$ [Eq.~(\ref{eq:phi-t})]. Instead of attempting to express $\pi(x)$ directly, we consider the net entropy production $\sigma(x)=\pi(x)-\phi(x)$. In steady state, $\int \sigma(x)\,dx=0$ [Eq.~(\ref{eq:entropybalance})], so $\sigma(x)$ is a zero-mean field. Motivated by the thermodynamic role of $T(x)$, we define the temperature deviation field $\theta(x)=T(x)-\langle T\rangle$, with $\langle \cdot \rangle$ denoting averaging with respect to $P(x)$. Rather than viewing $\sigma(x)$ as an arbitrary function of $x$, we regard it as a function of the thermodynamic field $\theta(x)$, and assume the form $P(x)f[\theta(x)]$, where $f$ is sufficiently smooth and can therefore be expanded in powers of $\theta(x)$. To enforce the zero-mean condition at each order, we introduce the mean-subtracted powers $u_n(x)\equiv \theta^n(x)-\langle \theta^n\rangle$, which differ from $\theta^n(x)$ only by a constant shift. Thus,
\begin{equation}
\sigma(x)=P(x)\sum_{n=1}^{\infty} c_n\,u_n(x).
\label{eq:unexpansion}
\end{equation}
Retaining only this leading order term, $u_1(x)=\theta(x)$, yields the inverse--Clausius closure
\begin{equation}
\sigma(x)=c\,\theta(x)\,P(x).
\label{eq:sigmatheta}
\end{equation}
The constant $c<0$ is fixed by matching the two sides at $x=0$ (see below). The thermodynamic interpretation is that $\sigma(x)$ is governed by temperature variations: The mean temperature separates hot and cold regions, and entropy transport reflects heat flowing from the former to the latter.

To complete the inverse--Clausius mapping, the constant $c$ is determined by matching Eq.~(\ref{eq:sigmatheta}) at a single point. We choose the symmetry point $x=0$, where evaluating $\pi(0)$ gives $\sigma(0)=\pi(0)-\phi(0)$, and thus $c=\sigma(0)/[P(0)\theta(0)]$. While this choice is natural, in practice the value of $c$ that best describes the data may differ slightly, depending on the region of interest. Since the mapping is formulated in terms of the reduced fields $P(x)$ and $\Delta(x)$, we express this evaluation in terms of them. In the two-state model, these fields fully determine the local state via $p_\pm=(P\pm\Delta)/2$, so that $\pi(x)$ can be written entirely in terms of $P$ and $\Delta$. In general multistate models this reduction is no longer exact, but we nevertheless proceed by expressing all quantities in terms of these fields alone.

To evaluate $\pi(0)$, we consider first the currents contribution in Eq.~(\ref{eq:kullback-leibler}). In the two-state model, 
where $p_\pm=(P\pm\Delta)/2$, 
the partial currents are given by
\begin{equation}
J_\pm(x)=(\mu F(x)\pm v_0)\,p_\pm(x)-D_0\,\partial_x p_\pm(x),
\end{equation}
which yields $J_+(x)-J_-(x)=v_0P(x)+\mu F(x)\Delta(x)-D_0\,\partial_x\Delta(x)$. At steady state, $J_+(x)=-J_-(x)=\big[v_0P(x)+\mu F(x)\Delta(x)-D_0\,\partial_x\Delta(x)\big]/2$. At $x=0$, symmetry implies that $F(0)=\Delta(0)=0$, so that $J_+(0)=\big[v_0P(0)-D_0\,\Delta'(0)\big]/2$, and the currents contribution then reads
\begin{equation}
\pi_{\mathrm{curr}}(0)=
\frac{\big[v_0P(0)-D_0\,\Delta'(0)\big]^2}{D_0\,P(0)}.
\label{eq:picurr}
\end{equation}
In the two-state model, the KL term vanishes at $x=0$, so $\pi(0)=\pi_{\mathrm{curr}}(0)$. In multistate models, however, this equality no longer holds. Nevertheless, $\pi_{\mathrm{curr}}(0)$, evaluated using the same expression in terms of $P$ and $\Delta$, is used as an approximation to the currents contribution.

We next consider the KL (second) term in Eq.~(\ref{eq:kullback-leibler}), associated with velocity reshuffling. This term is zero at $x=0$ in the two-state model, but is generally nonzero in multistate cases. Its evaluation requires the velocity distribution $P(0,v)$, which by symmetry satisfies $P(0,v)=P(0,-v)$. Within the reduced description, we base this estimate on the local second velocity moment $v_0^2$, which provides the natural measure of the strength of active force. Accordingly, we determine $P(0,v)$ by maximizing the relative entropy $-\int dv\,P(0,v)\ln[P(0,v)/P(v)]$ with respect to the reference velocity distribution $P(v)$, under the constraints of normalization and symmetry in $v$. This yields the Gaussian form: $P(0,v)/[P(0)P(v)] = e^{-\lambda v^2}\big/\!\int dv'\,P(v')e^{-\lambda v'^2}$,
where $\lambda$ is a Lagrange multiplier fixed numerically by the imposed value of $v_0^2$. In a four-state model, symmetry and normalization fix $P(0,v)$ uniquely via $v_0^2$, so that the maximum-entropy construction coincides with the exact velocity distribution at $x=0$, and the resulting KL term is exact. In general multistate models, this identification is no longer exact.

\begin{figure}[t]
\centering
\includegraphics[width=0.425\textwidth]{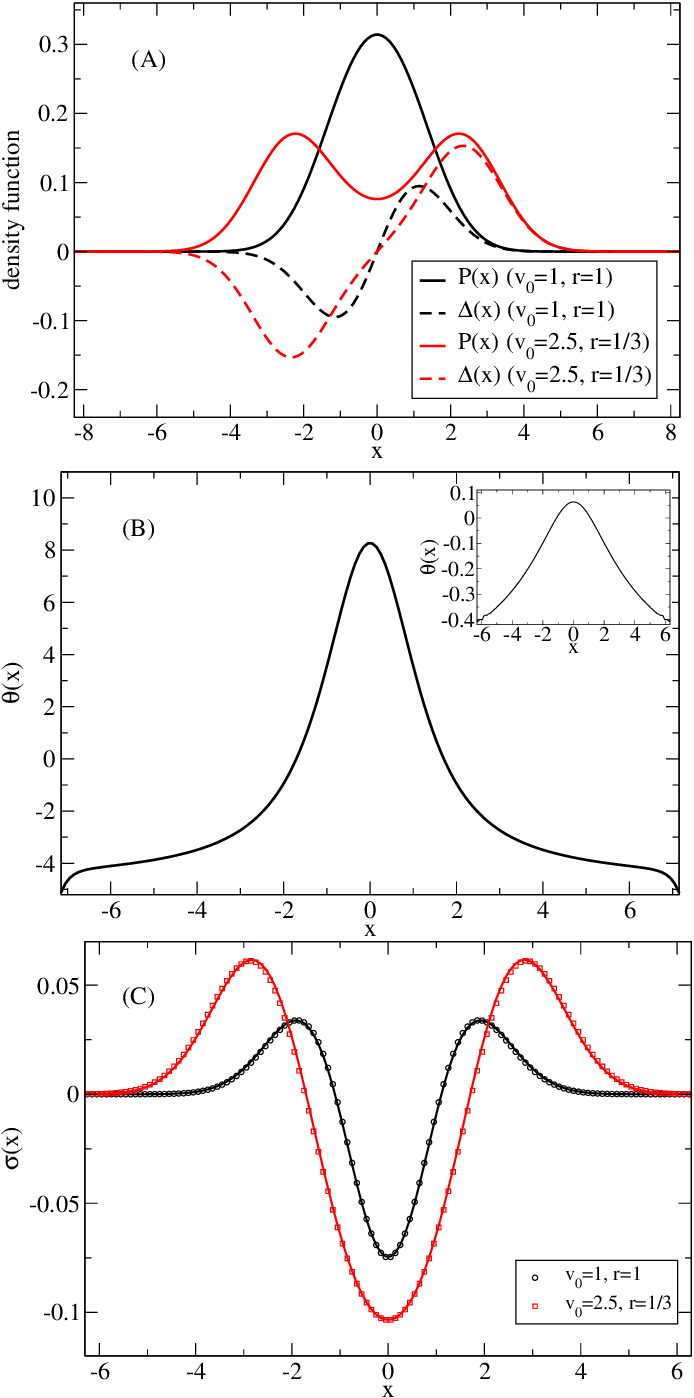}
\caption{
(A) Steady-state density $P(x)$ (solid curves) and normalized active flux density $\Delta(x)$ (dashed curves) for the s-RTP model in a harmonic potential. Black curves correspond to case (i), $v_0=1$, $r=1$, and red curves to case (ii), $v_0=2.5$, $r=1/3$. 
(B) Temperature field $\theta(x)$ obtained via the inverse--Clausius construction. Main panel: case (ii); inset: case (i).
(C) Net entropy production density $\sigma(x)$ for both cases. Symbols: exact results; solid curves: inverse-Clausius closure.}
\label{fig:p-delta}
\end{figure}

As a first example, we consider the s-RTP model in a harmonic potential $U(x)=x^2/2$ with $D=1$, and $\mu=1$. Figure~\ref{fig:p-delta}(A) shows simulation results for two parameter sets: (i) $v_0=1$, $r=1$, and (ii) $v_0=2.5$, $r=1/3$, chosen to yield qualitatively different steady-state densities $P(x)$. Case (i) is unimodal and close to equilibrium, whereas case (ii) is activity-dominated and bimodal. The active flux density $\Delta(x)$ has a similar structure in both cases, being negative for $x<0$, positive for $x>0$, and vanishing at large $|x|$. Figure~\ref{fig:p-delta}(B) shows the corresponding temperature deviation field $\theta(x)$, which exhibits pronounced spatial variations in the strongly active case (ii) (main panel), but remains small and nearly flat in case (i) (inset), consistent with the near-equilibrium form of $P(x)$. Figure~\ref{fig:p-delta}(C) shows the net entropy production density $\sigma(x)=\pi(x)-\phi(x)$, which in both cases is negative near the center, becomes positive at intermediate distances, and vanishes at large $|x|$, consistent with entropy flowing outward from a hotter central region. Symbols denote the exact (direct) evaluation, while solid lines show the inverse--Clausius closure, which is in perfect agreement with the direct results in both cases. 

As a second example, we consider the s-RTP model in a nonlinear confining potential $U(x)=\cosh(x)$, using the same two parameter sets as in Fig.~\ref{fig:p-delta}. Figure~\ref{fig:cosh} shows the net entropy production density $\sigma(x)$ for both cases, comparing the exact (symbols) with the inverse--Clausius closure (curves). In both cases, the agreement between the two is excellent across the entire domain. This demonstrates that the closure is not limited to harmonic confinement, but remains quantitatively accurate also for nonlinear potentials, supporting its robustness over a broad range of conditions.

\begin{figure}[t]
\centering
\includegraphics[width=0.425\textwidth]{fig2.eps}
\caption{
Net entropy production density $\sigma(x)$, as in Fig.~\ref{fig:p-delta}(C), but for the potential $U(x)=\cosh(x)$.}
 \label{fig:cosh}
\end{figure}

\begin{figure}[b]
\centering
\includegraphics[width=0.425\textwidth]{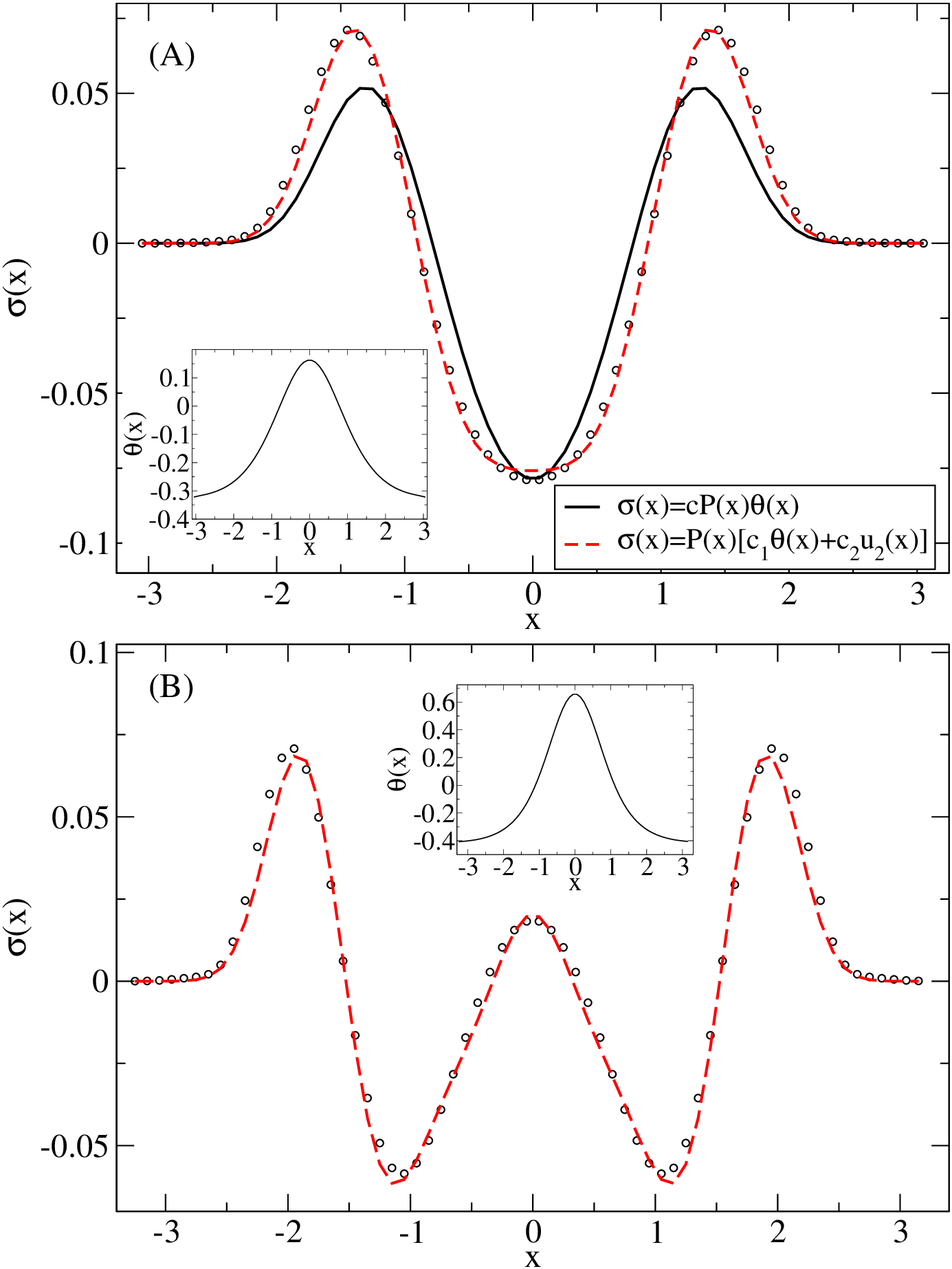}
\caption{
Net entropy production density $\sigma(x)$ for: (A) $U(x)=x^4/4$, (B) $U(x)=x^4/4-x^2$. In both cases, $r=1$ and $v_0=1$. Symbols denote the exact results. In panel (A), the black curve shows the leading-order inverse-Clausius closure [Eq.~(13)], while the red dashed curve is a fit including terms up to $u_2(x)$. In panel (B), the red dashed curve is a fit including terms up to $u_4(x)$.} The insets depict the corresponding temperature fields $\theta(x)$.
 \label{fig:quartic}
\end{figure}

As a third example, we consider the s-RTP model in a quartic confining potential, $U(x)=x^4/4$, with $v_0=1$ and $r=1$. Figure~\ref{fig:quartic}(A) shows the net entropy production density $\sigma(x)$ (circles), together with the inverse--Clausius prediction (black solid curve). In contrast to the previous cases, the linear form with $c$ fixed from the matching at $x=0$ no longer provides a good approximation. This deviation may be related to the spatial structure of $\sigma(x)$ near the origin, which is noticeably flatter in the quartic potential and appears to follow a similar shape to $U(x)$ in this region. In contrast, the temperature deviation field $\theta(x)$ (shown in the inset) retains a nearly quadratic profile. As a result, the relation $\sigma(x)\propto P(x)\theta(x)$ fails to capture the central behavior. This mismatch is resolved by including the next-order term, $u_2(x)=\theta^2(x)-\langle \theta^2\rangle$. A fit including this term yields very good agreement with the exact $\sigma(x)$ over the entire domain (red dahsed curve).

\begin{figure}[t]
\centering
\includegraphics[width=0.425\textwidth]{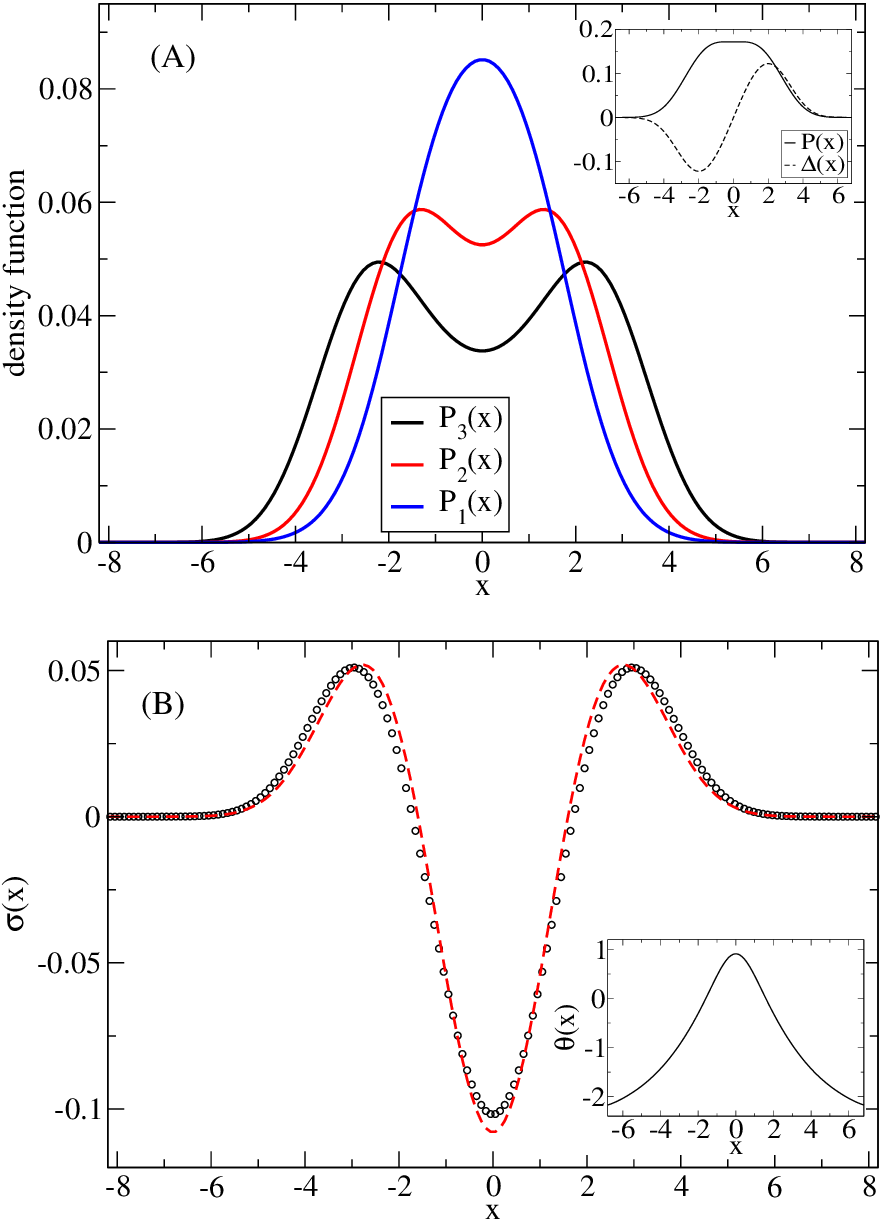}
\caption{(A) State-resolved SSDs $P_1(x)$, $P_2(x)$, and $P_3(x)$ for the six-state RTP in a harmonic trap with velocities $v=\pm1,\pm2,\pm3$. The inset shows the total density $P(x)=\sum_i P_i(x)$ and the corresponding AFD $\Delta(x)$, with $v_0\simeq1.87$. (B) Net entropy production $\sigma(x)$ with symbols denoting the direct evaluation and the curve the inverse-Clausius closure. The inset shows the corresponding temperature field $\theta(x)$.}
 \label{fig:6state}
\end{figure}

Panel (B) shows results for  $U(x)=x^4-2x^2$. In this case, $\sigma(x)$ takes a markedly different form in the central region, developing a clear local maximum at $x=0$. Interestingly, its shape again follows that of the potential, now reflecting the double-well structure, with a central maximum replacing the minimum observed in the previous cases. This suggests a broader trend in which the form of $\sigma(x)$ in the central region appears to mirror key features of the confining potential. The curve shows a fit including terms up to $u_4(x)$, which yields very good agreement with the exact results (circles), while lower-order truncations fail to reproduce the observed shape.

Finally, we consider a multistate RTP in the harmonic potential $U(x)=x^2/2$ with rate $r=2/3$, where at each tumbling event the velocity is redrawn from $v_i\in\{\pm1,\pm2,\pm3\}$ with equal probability. Figure~\ref{fig:6state}(A) shows the corresponding state-resolved stationary densities, which are clearly distinct. Their sum yields the total density $P(x)$ and the associated active flux density $\Delta(x)$ shown in the inset. In this model, the effective speed entering the inverse--Clausius construction is determined from the local second velocity moment, giving $v_0\simeq1.87$ for the present data. Figure~\ref{fig:6state}(B) compares the net entropy production density $\sigma(x)$ obtained from the inverse--Clausius construction (curve, with the corresponding temperature deviation field $\theta(x)$ shown in the inset) with the direct evaluation (symbols). To fix the closure constant $c$ in this multistate case, we follow the procedure described above to estimate $\pi(0)$, using an approximate evaluation of the currents contribution [expressed in terms of $P(x)$ and $\Delta(x)$] and the KL term. The agreement between the two is very good at the level of spatial structure, with small residual differences that are most pronounced near the origin. These reflect two distinct sources of error: The first arises from the approximations used in the origin estimate that fixes $c$, while the second originates from numerical differentiation in the direct evaluation of $\pi(x)$. Accordingly, the “exact” results are themselves subject to numerical error that accumulates as the number of velocity states increases.

To conclude, in this work we established an inverse--Clausius mapping between one-dimensional RTP dynamics and overdamped Brownian motion in a spatially inhomogeneous temperature field. The mapping combines an exact steady-state entropy flux $\phi(x)$, constructed from the polarization $\Delta(x)$, with a temperature field inferred from the steady-state density $P(x)$. This provides a thermodynamic description based solely on steady-state observables, without requiring explicit knowledge of the full position--velocity distribution $P(x,v)$. Within this framework, the net entropy production $\sigma(x)=\pi(x)-\phi(x)$ is related to the temperature fluctuation field $\theta(x)$, leading to the linear closure $\sigma(x)=c\,P(x)\theta(x)$. This formulation provides a physical picture in which entropy production is governed by an emergent temperature landscape, with entropy transported from locally hotter to colder regions. The closure is practically exact for the two-state model under harmonic confinement and for nonlinear potentials with positive curvature at the origin. In multistate models, it remains accurate, with only small deviations. In contrast, potentials with vanishing or negative curvature at the origin require systematic extensions through higher-order terms. In these cases, the spatial structure of $\sigma(x)$ in the central region tracks the underlying potential, transitioning from a single-well to a double-well form. These results indicate that the inverse--Clausius framework captures key features of entropy production, while highlighting the role of the confining potential in shaping its spatial organization.

\bibliography{apssamp}

\end{document}